\documentclass[runningheads]{llncs}
\usepackage{graphicx}
\usepackage{siunitx}
\usepackage{graphicx,subfigure}
\usepackage{tikz}
\usepackage{comment}
\usepackage{amsmath,amssymb} % define this before the line numbering.
\usepackage{color}
\usepackage{multirow}
\usepackage{makecell}
\usepackage{booktabs}
\usepackage{adjustbox}
\usepackage[accsupp]{axessibility}  % Improves PDF readability for those with disabilities.

\usepackage[colorlinks, linkcolor=red, anchorcolor=yellow, citecolor=green]{hyperref}

\newcommand\blfootnote[1]{%
	\begingroup
	\renewcommand\thefootnote{}\footnote{#1}%
	\addtocounter{footnote}{-1}%
	\endgroup
}

\begin{document}
% \linenumbers
\pagestyle{headings}
\mainmatter
\def\ECCVSubNumber{11}  % Insert your submission number here

\title{Efficient Image Super-Resolution using Vast-Receptive-Field Attention} % Replace with your title

%\begin{comment}
% \titlerunning{ECCV-22 submission ID \ECCVSubNumber} 
% \authorrunning{ECCV-22 submission ID \ECCVSubNumber} 
% \author{Anonymous ECCV submission}
% \institute{Paper ID \ECCVSubNumber}
%\end{comment}
%******************

% \begin{comment}
% \author{Lin Zhou\inst{1}\orcidID{} \and
\author{Lin Zhou\inst{1,*} \and
Haoming Cai\inst{1,*} \and
Jinjin Gu\inst{2,3} \and
Zheyuan Li\inst{1} \and
Yingqi Liu\inst{1},\\
Xiangyu Chen\inst{1,2,4} \and
Yu Qiao\inst{1,2} \and
Chao Dong\inst{1,2,}$^\dagger$}
\authorrunning{L. Zhou et al.}
\institute{ShenZhen Key Lab of Computer Vision and Pattern Recognition, SIAT-SenseTime Joint Lab, Shenzhen Institutes of Advanced Technology, Chinese Academy of Sciences \and
Shanghai AI Laboratory, Shanghai, China \and
The University of Sydney \quad\quad \inst{4} University of Macau \\
\email{\{zhougrace885, helmut.choy, chxy95\}@gmail.com, jinjin.gu@sydney.edu.au}\\
\email{\{zy.li3, yq.liu3, yu.qiao, chao.dong\}@siat.ac.cn}
}
% \end{comment}
%******************
\maketitle

\begin{abstract}
The attention mechanism plays a pivotal role in designing advanced super-resolution (SR) networks. In this work, we design an efficient SR network by improving the attention mechanism. We start from a simple pixel attention module and gradually modify it to achieve better super-resolution performance with reduced parameters. The specific approaches include: (1) increasing the receptive field of the attention branch, (2) replacing large dense convolution kernels with depth-wise separable convolutions, and (3) introducing pixel normalization. These approaches paint a clear evolutionary roadmap for the design of attention mechanisms. Based on these observations, we propose VapSR, the VAst-receptive-field Pixel attention network. Experiments demonstrate the superior performance of VapSR. VapSR outperforms the present lightweight networks with even fewer parameters. And the light version of VapSR can use only 21.68\% and 28.18\% parameters of IMDB and RFDN to achieve similar performances to those networks. The code and models are available at \url{https://github.com/zhoumumu/VapSR}.
\blfootnote{* Equal Contributions, $\dagger$ Corresponding Author.}
\end{abstract}

\section{Introduction}
Single image Super-Resolution (SISR) is a fundamental low-level vision problem that aims at recovering a high-resolution (HR) image from its low-resolution (LR) observations.
SISR has attracted increasing attention in both the research community and industry.
Since SRCNN \cite{dong2014learning} introduced deep learning into SR, deep networks have become the de facto approach for advanced SR algorithms due to their ease of use and high performance.
However, deep SR networks rely on a large number of parameters that can provide sufficiently complex capacity to map LR images to HR images.
These parameters and high computation costs limit the application of SR networks.
The design of SR networks with efficiency as the primary goal has gradually become an important issue.

Among the numerous SR networks, the studies related to the attention mechanism have achieved a lot of success.
The channel attention brought by RCAN \cite{zhang2018image} makes it practical to train very deep high-performance SR networks.
PAN \cite{zhao2020efficient} has achieved good progress in designing a lightweight SR network using pixel attention.
After image processing entered the Transformer era, the application of the attention mechanism underwent great changes.
Vision Transformers \cite{vit} rely on attention mechanisms to achieve excellent performance.
Many works have proved that introducing large receptive fields and local windows \cite{chen2022activating,shi2022rethinking} in the attention branch improves the SR effect.
However, many advanced design ideas have not been verified in designing the attention mechanism for convolutional lightweight SR networks.
In this paper, we start from a basic pixel attention module to explore better attention mechanisms designed for efficient SR.

The first effort we made in this paper was to introduce the large receptive field design into the attention mechanism.
This is in line with other recent design trends using large kernel sizes \cite{guo2022visual}, as well as the design principles of transformers \cite{chen2022activating,liang2021swinir,shi2022rethinking}.
We show the advantages of using large kernel convolutions in the attention branch.
Secondly, we use depth-wise separable convolution to split dense large convolution kernels.
A large receptive field is achieved in the attention branch using a depth-wise and a depth-wise dilated convolution.
We also replace the $3\times3$ convolutions in the backbone network with $1\times1$ convolutions to reduce the number of parameters.
Thirdly, we present a novel pixel normalization that can make the training less prone to crashing.

Along the above footprints, we demonstrate a novel path to an efficient SR architecture called VapSR (VAst-receptive-field Pixel attention network).
Compared with the current state-of-the-art algorithms, the proposed VapSR reduces a lot of parameters while improving the SR effect. 
For example, compared with the champion of the NTIRE2022 efficient SR competition \cite{kong2022residual}, VapSR achieves an improvement on PSNR by more than 0.1dB with 185K fewer parameters.
Our experiments demonstrate the effectiveness of the proposed method.

\section{Related Work}

\textbf{Deep Networks for SR.}
Since SRCNN \cite{dong2014learning} was proposed as the pioneering work for employing a three-layer convolutional neural network for the SR task, numerous methods \cite{dong2016accelerating,kim2016accurate,ledig2017photo,lim2017enhanced,tai2017image,qian2019rethinking} have been proposed to achieve better performance. 
FSRCNN \cite{dong2016accelerating} proposes a pipeline that upsamples features at the end of the network, which boosts the performance while keeping the model lightweight.
VDSR \cite{kim2016accurate} introduces skip connections for residual learning to increase the depth of the SR network.
DRCN \cite{kim2016deeply} and DRRN \cite{tai2017image} both adopt the recursive structure to improve the reconstruction performance.
SRDenseNet \cite{tong2017image} and RDN \cite{zhang2018residual} prove that the dense connection is beneficial to improving the capacity of SR models. 
RCAN \cite{zhang2018image} employs a channel attention scheme to bring the attention mechanism to the SR methods.
SAN \cite{dai2019second} proposes a second-order attention module, which brings further performance improvement.
SwinIR \cite{liang2021swinir} promotes Swin Transformer \cite{swin_t} for the SR task.
HAT \cite{chen2022activating} refreshes state-of-the-art performance through hybrid attention schemes and pre-training strategy.

\textbf{Attention Schemes for SR.}
The attention mechanism can be interpreted as a way to bias the allocation of available resources towards the most informative parts of an input signal. 
There are approximately four attention schemes: channel attention, spatial attention, combined channel-spatial attention and self-attention mechanism. 
RCAN \cite{zhang2018image}, inspired by SENet \cite{hu2018squeeze}, reweights the channel-wise features according to their respective weight responses.
SelNet \cite{choi2017deep} and PAN \cite{zhao2020efficient} employ the spatial attention mechanism, which calculates the weight for each element. 
Regarding combined channel-spatial attention, HAN \cite{niu2020single} additionally proposes a spatial attention module via 3D convolutional operation.
The self-attention mechanism was adopted from natural language processing to model the long-range dependence \cite{transformer}.
IPT \cite{chen2021pre} is the first Transformer-based SR method based on the ViT \cite{vit}. It relies on a large model scale (over 115.5M parameters).
SwinIR \cite{liang2021swinir} calculates the window-based self-attention to save the computations.
HAT \cite{chen2022activating} further proposes multiple attention schemes to improve the window-based self-attention and introduce channel-wise attention to SR Transformer.

\textbf{Efficient SR Models.}
Efficient SR designing aims to reducing model complexity and latency for SR networks \cite{ahn2018fast,hui2018fast,hui2019lightweight,RFDN,Kong_2022_CVPR,li2022blueprint,chen2021attention}. 
CARN \cite{ahn2018fast} employs the combination of group convolution and $1\times1$ convolution to save computations.
After IDN \cite{hui2018fast} proposed the residual feature distillation structure, there appears a series of works \cite{hui2019lightweight,RFDN,li2022blueprint} following this micro-architecture design.
IMDN \cite{hui2019lightweight} improves IDN via an information multi-distillation block by using a channel splitting strategy.
RFDN \cite{RFDN} rethinks the channel splitting operation and introduces the progressive refinement module as an equivalent architecture.
In NTIRE 2022 Efficient SR Challenge \cite{li2022ntire}, RLFN \cite{Kong_2022_CVPR} won the championship in the runtime track by ditching the multi-branch design of RFDN and introducing a contrastive loss for faster computation and better performance.
%
% The main idea is to use three convolutional layers for residual local feature learning to simplify feature aggregation, which achieves a good trade-off between model performance and inference time.
%
BSRN \cite{li2022blueprint} won the first place in the model complexity track by replacing the standard convolution with a well-designed depth-wise separable convolution to save computations and utilizing two effective attention schemes to enhance the model ability. 

\textbf{Large Kernel Design.}
CNNs used to be the common choice for computer vision tasks. However, CNNs have been greatly challenged by Transformers recently \cite{vit,detr,swin_t,uniformer}, and Transformer-based methods have also shown leading performances on the SR task \cite{chen2021pre,liang2021swinir,edt,chen2022activating}.
In Transformer, self-attention is designed to be either global \cite{vit,chen2021pre} or local, both accompanied by larger kernels \cite{swin_t,liang2021swinir,chen2022activating}. Thus, information can be gathered from a large region.
Inspired by this characteristic of Transformer, a series of works have been proposed to design better CNNs \cite{convmixer,liu2022convnet,ding2022scaling,guo2022visual}.
ConvMixer \cite{convmixer} utilizes large kernel convolutions to build the model and achieve the competitive performance to the ViT \cite{vit}.
ConvNeXt \cite{liu2022convnet} proves that well-designed CNN with large kernel convolution can obtain similar performance to Swin Transformer \cite{swin_t}.
RepLKNet \cite{ding2022scaling} scales up the filter kernel size to $31\times31$ and outperforms the state-of-the-art Transformer-based methods.
VAN \cite{guo2022visual} conducts an analysis of the visual attention and proposes the large kernel attention based on the depth-wise convolution.

\begin{figure}[!tbp]
    \centering
    \includegraphics[width=1\textwidth]{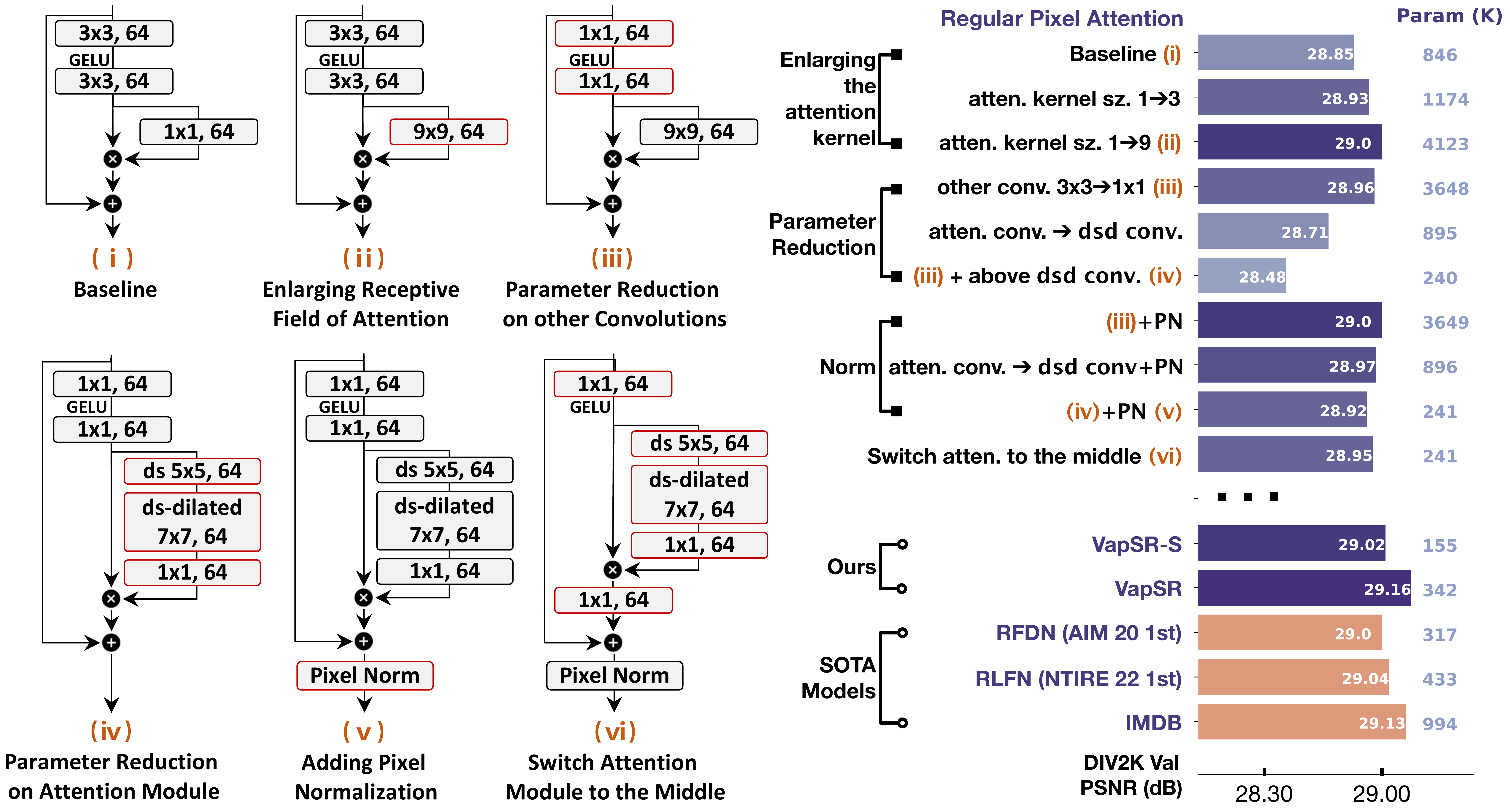}
    \caption{The evolutionary design roadmap of the proposed method. The figures on the left are the key architectural milestones. The plot on the right shows the main models' parameters and PSNR performance on DIV2K validation set. Every evolution and modification of the main design stages are marked with \textcolor{red}{red} boxes on the left and described with the text on the right. We omit some micro designs in this plot and they will be elaborated lately in section \ref{sec:ablation}.
    }
    \label{fig:roadmap}
\end{figure}

\section{Motivation}
\label{sec:3.1}
The attention mechanism has been proven effective in SR networks.
In particular, an efficient SR model PAN \cite{zhao2020efficient} achieves good performance using pixel attention while greatly reduces the number of parameters.
Pixel attention performs an attention operation on each element of the features.
Compared with channel attention and spatial attention, pixel attention is a more general form of attention operation and thus provides a good baseline for our further exploration.

Inspired by recent advances in self-attention \cite{transformer} and vision transformers \cite{vit}, we believe that there is still considerable room for improvement even for the attention mechanism based on convolutional operations.
In this section, we show the process of improving SR network attention through three design criteria in pixel attention.
First, we show the advantages of using large kernel convolutions in the attention branch.
Then we use well-designed depth-wise separable large kernel convolutions to reduce the huge computational burden brought by large kernel convolutions.
We demonstrate the potential of this network topology design for efficient SR.
Finally, inspired by vision Transformers, we introduce a pixel-wise normalization operation in the convolutional network to train SR networks with complex attention efficiently and stably.
We demonstrate a solid roadmap toward our improved network design.

\subsection{Large Kernel in Visual Attention}
We start with the building block of one of the most common SR networks with a basic pixel attention operation.
This block is shown in \figurename~\ref{fig:roadmap} (i).
In general, the main operation that provides receptive fields in SR networks are the two $3\times3$ convolutions, whereas the attention branch contains only one $1\times1$ convolution.
However, inspired by the vision Transformers, we may improve the performance by increasing the receptive field of the attention branch.
We enlarge the kernel size in the attention branch of the baseline block to 3 and 9 to study the effect of enlarging the attention kernel, respectively.
This modification is shown in \figurename~\ref{fig:roadmap} (ii).
The performance variation is shown by the first 3 experiments in the roadmap on the right of \figurename~\ref{fig:roadmap}.
It can be seen that although it brings tons of additional parameters, enlarging the kernel size in attention brings about 0.15dB of performance improvement.
After showing that performing large kernel convolutions in attention can bring benefits, we continue to explore this basis.

\begin{figure}[t]
\begin{minipage}[t]{0.63\textwidth}
\centering
    \includegraphics[width=0.85\linewidth]{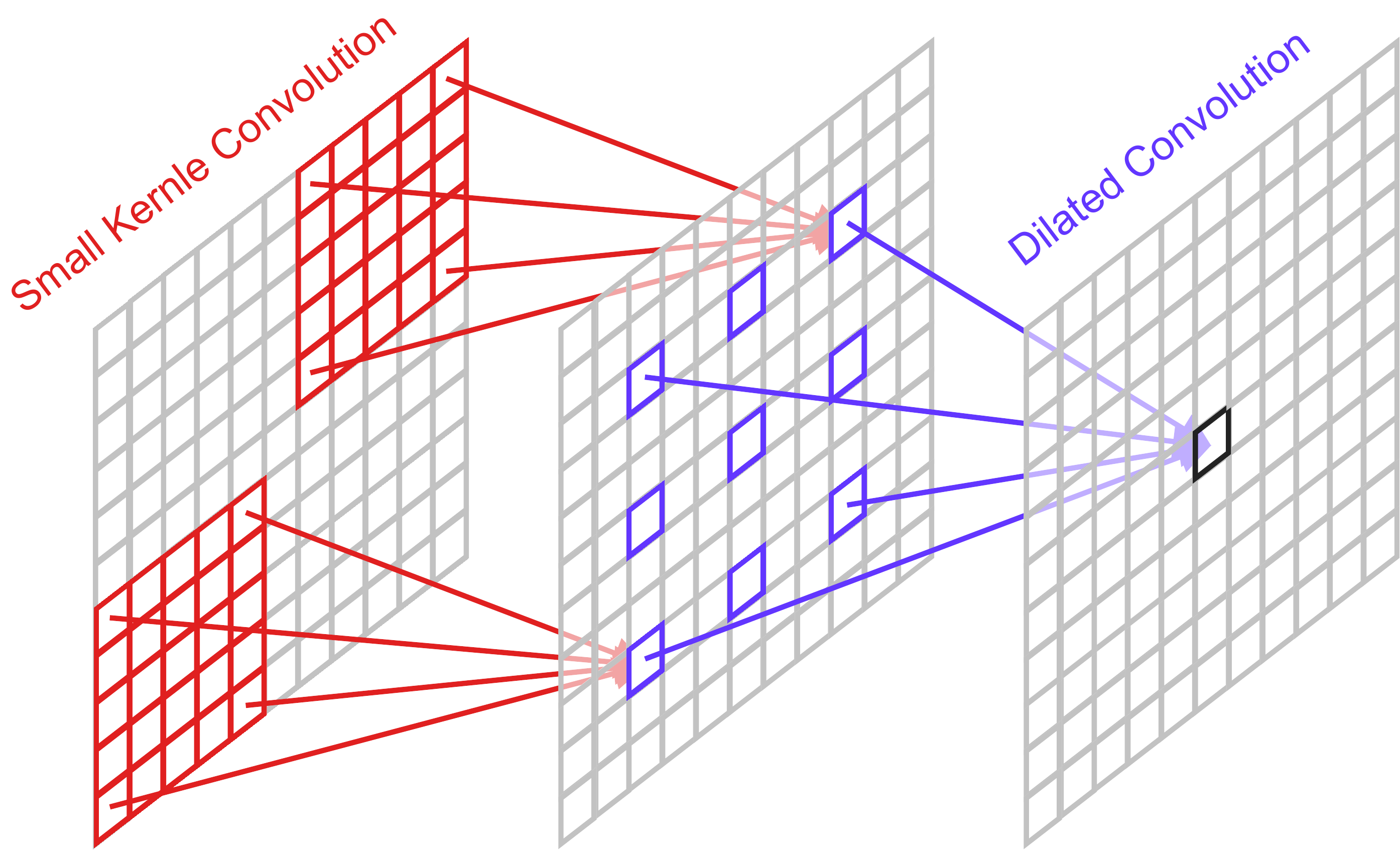}
    \caption{
    An $11\times11$ receptive field can be replaced by a $5\times5$ small convolution and a $3\times3$ dilated convolution with a dilation of 3. This operation saves the number of parameters and achieves a large receptive field.}
    \label{fig:depthwise}
\end{minipage}
\hfill
\begin{minipage}[t]{0.34\textwidth}
    \centering
    \includegraphics[width=\linewidth]{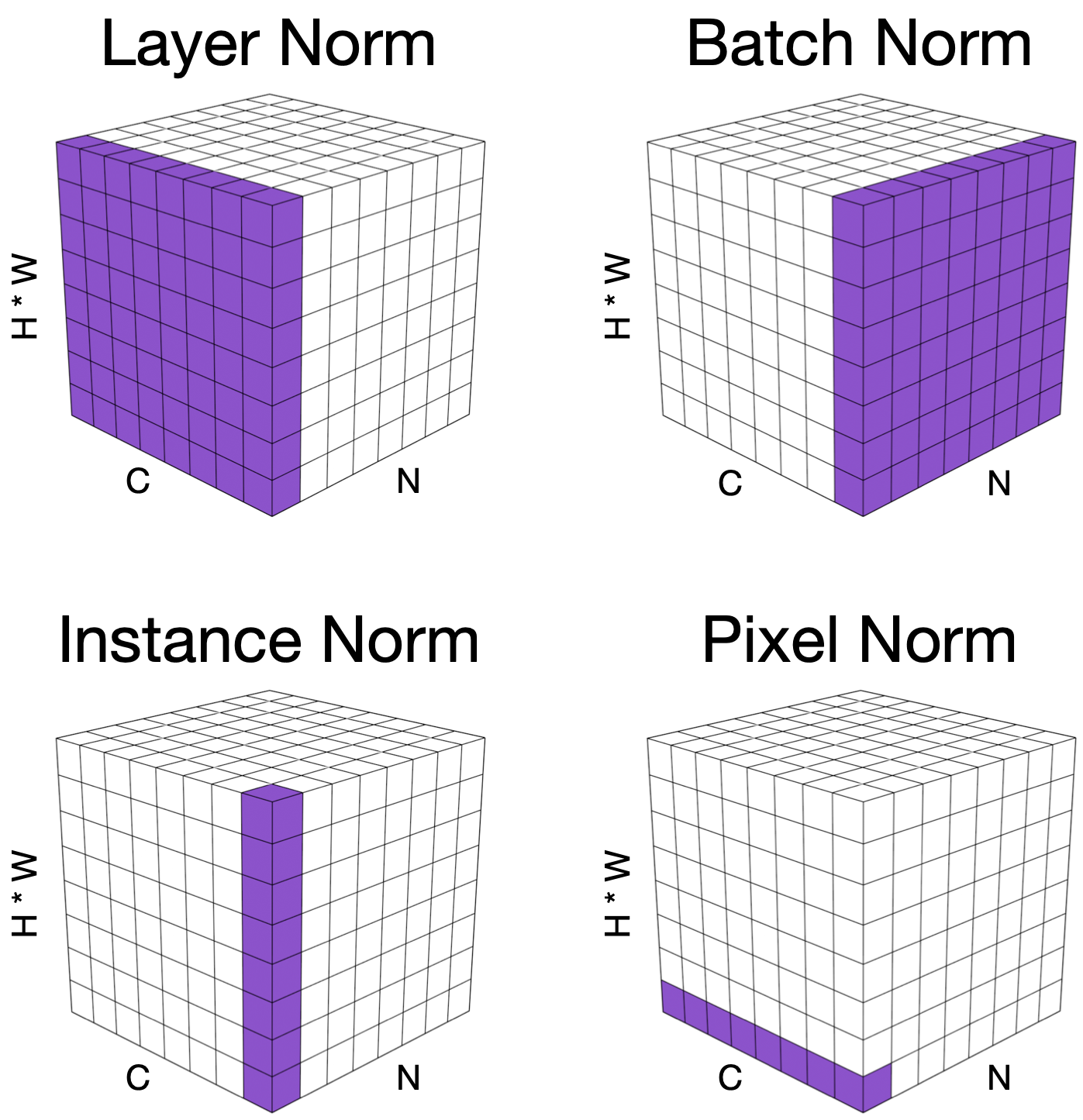}
    \caption{Different normalization methods. The cubes in purple are normalized by the same mean and variance.}
    \label{fig:norm}
\end{minipage}
\end{figure}

\subsection{Parameter Reduction.}
The above kernel enlarging strategy provides an architecture that relies on one large kernel convolution in attention and two $3\times3$ convolutions outside the attention.
It brings a large number of parameters, so we try to remove relatively unimportant parts of the network as much as possible and reduce parameters accordingly.
The good news is that dense convolution kernels are often not the best choice for large kernel size.
We can reduce the parameters of the network by implementing a more sparse large kernel convolution.

Depth-wise separable convolution is a classic, intuitive solution for large kernel convolution parameter reduction.
Depth-wise separable convolution splits a dense convolution operation into spacial depth-wise convolutions and pointwise convolutions in channels.
The depth-wise convolution is in the form of a group convolution that assigns only one kernel for each feature channel.
The pointwise convolution is a $1\times1$ convolution for channel fusion.
Inspired by \cite{guo2022visual}, the depth-wise convolution can be further decomposed.
Taking a convolution of size $11\times11$ as an example, we can convert it into a $5\times5$ normal convolution and a $3\times3$ dilated convolution with a dilation of 3 while keeping its equivalent receptive field size unchanged, as shown in \figurename~\ref{fig:depthwise}.
This design reduces the number of parameters as much as possible while making the receptive field even larger.
The effect is shown in the fifth experiment on the right of \figurename~\ref{fig:roadmap}. A performance drop is observed compared with the third experiment, however the replacement saves about 3,200K parameters.
In addition to the above solution, compared with the large kernel attention operation, the receptive field brought by the rest $3\times3$ convolutions in the original backbone is no longer important.
We replace the two $3\times3$ convolutions with $1\times1$ convolutions and find more parameter reduction of 655K.
%
% However, the performance drops dramatically after this modification.
%
% By combining the depth-wise separable convolution attention and the $1\times1$ body convolutions, we reduce the parameter to 241K. The implementation of such architecture is shown in \figurename~\ref{fig:roadmap} (iv).
By combining the depth-wise separable convolution attention and the $1\times1$ body convolution, as shown in \figurename~\ref{fig:roadmap} (iv), we compress the model size to the limit.
We argue that with good training of this network, we might be able to achieve high performance with such little parameters.

\subsection{Pixel Normalization for Stable Attention Training}
Due to the introduction of element-wise multiplication in the attention mechanism, the training stability is greatly reduced.
At a small learning rate, the network cannot converge well, but increasing the learning rate will cause the network returning abnormal gradients.%, and training collapse.
The above parameter reduction solution produces such a difficult-to-train network that it suffers from performance drop. %of about 0.5dB.

We find that this training problem is partly due to internal covariate shift \cite{ioffe2015batch} phenomena.
For a network with attention layers, its multiplication makes the degree of shift more difficult to control.
To solve this problem, we introduce a pixel normalization layer to normalize the shifted layer distribution to a standard normal distribution.
The difference between pixel normalization and other normalization methods is shown in \figurename~\ref{fig:norm}.
Given a feature tensor that can be formulated as $x\in\mathbb{R}^{HW\times C}$, ($H$, $W$ and $C$ are the height, width and feature dimension), $x$ can be viewed as $HW$ feature vectors, and each vector belongs to a pixel position.
We represent the feature vector of the $i$th pixel with $x^i\in\mathbb{R}^{C}$.
The mean and variance of $x^i$ are:
\begin{equation}
    \mu^i=\frac{1}{C}\sum_{j=1}^{C} x^i_j, \quad \sigma^i=\frac{1}{C}\sum_{j=1}^C (x_j^i-\mu^i)^2.
\end{equation}
The output of the pixel normalization can be formulated as
\begin{equation}
    \tilde{x}^i=\frac{x^i-\mu^i}{\sqrt{\sigma^i + \epsilon}}\odot\gamma+\beta,
\end{equation}
where $\gamma$ and $\beta$ represent the parameter vectors for scaling and shifting. They have the same dimensions as $C$.
Different from the other normalization methods in the existing literature, pixel normalization calculates the mean and variance of the features of different pixels and normalizes them separately.
In other words, pixel normalization's shifting and scaling operations are spatially inhomogeneous.

The boost from using pixel normalization is huge.
According to \figurename~\ref{fig:roadmap}, using pixel normalization on the reduced parameter model yields excellent results close to that of using a large dense kernel in attention.
Equipped with the above two practices, the network's parameters reduce to 241K on the basis of outperforming the baseline.
Now we get the novel architecture shown in \figurename~\ref{fig:roadmap} (v).

\subsection{Discussion}
The generated network design correlates with some existing models in several respects.
Firstly, in vision transformers, layer normalization has been proved important and effective \cite{vit,liang2021swinir}.%proven important for the good performance \cite{vit,liang2021swinir}.
The Transformer networks \cite{liang2021swinir} usually reshape the feature map from $C\times H\times W$ to $HW\times C$ and then perform layer normalization.
However, it's no longer consistent with the layer normalization originally used for convolution networks \cite{ba2016layer} at this time, but is equivalent to the pixel norm described in this paper when the token size is $1\times1$.
The original layer normalization will introduce parameter numbers consistent with the element number in the feature tensor. Thus, models built with it can no longer handle arbitrary resolutions.
The pixel normalization we describe is more flexible and efficient than many other known normalization methods.
Note that although the equivalent method of our pixel normalization has already appeared in the vision transformers, its successful application in SR task has not yet been witnessed.

We also found that our findings are very similar to a concurrent work, the large kernel attention (LKA) \cite{guo2022visual}.
One main difference is that LKA uses two $1\times1$ convolutional layers as projection layers and places the attention layer in the middle of the projection layers.
This design is also very similar to Transformer's use of the attention mechanism.
\figurename~\ref{fig:roadmap} shows the result when using a similar approach to LKA.
It can be seen that changing layer order brings a performance improvement of 0.03dB.
At this point, our method is shown in the \figurename~\ref{fig:roadmap} (vi).
We then make more micro designs and refinement on this building block which enumerated in section \ref{sec:ablation}, forming into the eventual proposed architecture.
We proved that our proposed method achieves SR performance higher than the existing methods with fewer parameters.

\begin{figure}[!tbp]
    \centering
    \includegraphics[width=1.0\textwidth]{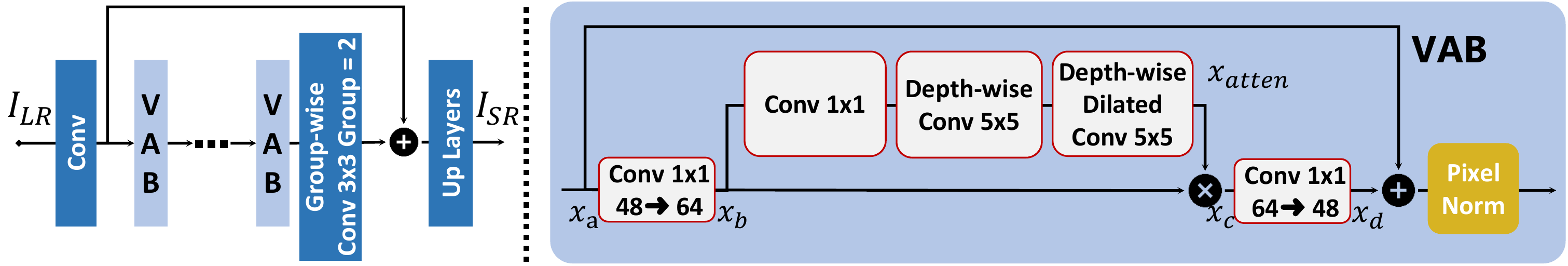}
    \caption{The architecture of the proposed VapSR. The block on right is the detailed illustration of the proposed main block VAB. $I_{LR}$ and $I_{SR}$ are the corresponding input low-resolution image and the output super-resolution image of the network.}
    %Up Layer is the sequential combination of convolution layer, $\times$2  pixelshuffle, convolution layer, and $\times$2 pixelshuffle.
    \label{fig:network}
\end{figure}

\section{Network Architecture}
Based on the building block discussed above, we build a novel SR network called VapSR (VAst-receptive-field Pixel attention network).
The architecture is illustrated in \figurename~\ref{fig:network}.
The high-level design of the proposed network follows the common design of deep SR networks.
VapSR contains three modules: (1) feature extraction, (2) nonlinear mapping, and (3) reconstruction. 

Given a low-resolution image $I_{LR}$, the feature extraction contains a convolution layer with a kernel size of $3\times3$ to extract features from $I_{LR}$.
\begin{equation}
    x_0 = f_{ext}(I_{LR}),
\end{equation}
and $x_0$ is the extracted higher dimensional feature maps.

In the nonlinear mapping stage, $x_0$ is fed into a stack of the building blocks to enhance the feature representations. 
We denote the building block as $f_{VAB}$(·), and this process can be formulated as
\begin{equation}
    x_n = f^n_{VAB}(f^{n-1}_{VAB}(...f^0_{VAB}(x_0)...)),
\end{equation}
where $x_n$ represents the output feature map of the $n$th VAB. 
At the end of the nonlinear mapping stage, we add a $3\times3$ convolution layer $f_{ref}$(·) after the building blocks and perform a residual connection with $x_0$:
\begin{equation}
    x_{map}=f_{ref}(x_n) + x_0.
\end{equation}

At last, we utilize the reconstruction module to upsample the features to the HR size. Here we obtain:
\begin{equation}
    I_{SR} = f_{rec}(x_{map}),
\end{equation}
where $f_{rec}$(·) denotes the reconstruction module, and $I_{SR}$ is the final result of the network.
Our reconstruction module contains two $\times2$ pixel-shuffle layers to implement $\times4$ upsampling scale, which brings a consistent promotion compared to a single $\times4$ pixel-shuffle layer.
There are convolution layers before both pixel-shuffle layers, and the number of channels can be adjusted as required.

\subsection{The Building Blocks}
\label{sec:methodVAB}
The building block is generally modified from \figurename~\ref{fig:roadmap} (vi) and is shown in \figurename~\ref{fig:network}.
As described above, in each block we have two regular $1\times1$ convolution layers, and a depth-wise separable large kernel attention in the middle.
We also have a pixel normalization at the end of each block.

Given input feature $x_a$, the first $1\times1$ convolution layer projects $x_a$ to $x_b$ and expands the number of channels from 48 to 64.
We perform GELU activation\cite{hendrycks2016gaussian} to $x_b$, and this is the only activation in the building block.
In the attention module, we firstly use a $1\times1$ pointwise convolution for channel fusion. Then we use a depth-wise convolution with a kernel size of $5$ and a depth-wise dilation convolution with a kernel size of $5$ and dilation of $3$. The combination of these two convolution layers is able to implement a receptive field of $17$.
The feature $x_{atten}$ generated by the attention branch is the same size as the original feature $x_b$ through reasonable padding.
The attention is implemented using an element-wise product as $x_c=x_{atten}\odot x_b$.
Then, another $1\times1$ convolution layer projects $x_c$ to $x_d$ and shrinks the number of channels back to 48.
At last, the pixel normalization is performed on $x_d+x_a$.

% %
% Illustration of bottleneck and $conv_{body}$ can be found in \ref{method3.2}
% %
% With these efficient designs, the last evolution finally reaches the proposed VAB. 
% %
% By VAB and other micro designs, we construct a tiny version of VapSR, named VapSR-S. 
% %
% Section \ref{exp5.1} explains the difference between these two models.
% %
% In the last step of the road map, we train the VapSR-S with the popular experimental setup used for NTIRE 2022 Efficient SR Challenge.
% %
% As shown in \ref{fig:roadmap}, the proposed VapSR-S achieves comparable performance with less model complexity compared to direct competitors, such as PAN, BSRN-S, and RFDN.

\section{Experiments}

\subsection{Experimental Setup}
\label{exp5.1}

\textbf{Datasets and Evaluation Metrics.}
The training images consist of 2650 images from Flickr2K \cite{lim2017enhanced} and 800 images from DIV2K \cite{agustsson2017ntire} train. 
We evaluate our models on widely used benchmark datasets: Set5 \cite{bevilacqua2012low}, Set14 \cite{zeyde2010single}, BSD100 \cite{martin2001database}, and Urban100 \cite{huang2015single}. 
The commonly used data augmentation methods are applied in the training dataset.
Specifically, We use the random combination of random rotation of \ang{0}, \ang{90}, \ang{180}, \ang{270} and horizontal flipping for data augmentation.
The average peak-signal-to-noise ratio (PSNR) and the structural similarity \cite{wang2004image,jinjin2020pipal} (SSIM) on the luminance (Y) channel are used as the evaluation metrics.

\textbf{Implementation Details.}
We implement two models, VapSR and VapSR-S. VapSR $\times4$ consists of 21 VABs and VapSR-S (also for the $\times4$ scale) is the light version of VapSR with 11 VABs. And we configure the input and output feature to 32 channels instead of 48 channels for VapSR-S. Both of them have two $\times2$ pixel-shuffle layers and two convolution layers. We make minor adjustment on the Up-Layers and the number of blocks for $\times 2$ and $\times 3$ scale.
%
% We implement two models, VapSR and VapSR-S, and four different configurations in total. 
% %
% VapSR $\times4$ consists of 21 VABs and VapSR-S (also for the $\times4$ scale) is the light version of VapSR with 11 VABs. And we configure the input and output feature to 32 channels instead of 48 channels for VapSR-S.
% %
% For the upsampling layers in reconstruction modules, both VapSR and VapSR-S have two $\times2$ pixel-shuffle layers and two convolution layers.
% %
% Besides, VapSR for $\times 2$ and $\times 3$ scale use 20 VABs, and adopt a single-layer pixel-shuffle module and still two convolution layers to accomplish reconstruction.

\textbf{Training Details.}
The model is trained using the Adam optimizer \cite{kingma2014adam} with $\beta_1= 0.9$ and $\beta_2 = 0.99$. Notably, using $\beta_2 = 0.99$ stead of commonly used $\beta_2$ = 0.999 can bring better performance for our proposed model design. The learning rate is set to $1\times10^{-3}$ during the whole $1 \times 10^6$ training iterations. And we set a smaller learning rate specially for the $\times 2$ scale. The weight of the exponential moving average (EMA) \cite{athiwaratkun2018there} is set to 0.999. Only the L1 loss is used to optimize the model.
For VapSR, the mini-batch and the input patch size are set to 64 and $48\times48$. We enlarge the setting to 192 and $64\times64$ for VapSR-S.

\begin{table*}[!htbp]
\centering
\caption{Quantitative comparison with state-of-the-art methods on benchmark datasets. The best and second-best performance are in \textcolor{red}{red} and \textcolor{blue}{blue} colors respectively. 'Multi-Adds' is calculated with a $1280\times720$ GT image.}
\label{table:comparison_SOTA}
\resizebox{\linewidth}{!}{
\begin{tabular}{l|c|cc|ll|ll|ll|ll}
    \toprule
    \multicolumn{1}{c|}{\multirow{2}{*}{Method}} & \multirow{2}{*}{Scale} & \multirow{2}{*}{Params[K]} & \multirow{2}{*}{Multi-Adds[G]} & \multicolumn{2}{c|}{Set5} & \multicolumn{2}{c|}{Set14} & \multicolumn{2}{c|}{B100} & \multicolumn{2}{c}{Urban100} \\ %\cline{5-12} 
    \multicolumn{1}{c|}{} & & & & \multicolumn{2}{l|}{PSNR/SSIM} & \multicolumn{2}{l|}{PSNR/SSIM} & \multicolumn{2}{l|}{PSNR/SSIM} & \multicolumn{2}{l}{PSNR/SSIM} \\ \midrule
    
    Bicubic & \multirow{16}{*}{$\times 2$} &  -   &           -           & \multicolumn{2}{l|}{33.66/0.9299} & \multicolumn{2}{l|}{30.24/0.8688} & \multicolumn{2}{l|}{29.56/0.8431} & \multicolumn{2}{l}{26.88/0.8403} \\
    SRCNN \cite{dong2014learning}                  &                        & 8        &  52.7      & \multicolumn{2}{l|}{36.66/0.9542} & \multicolumn{2}{l|}{32.45/0.9067} & \multicolumn{2}{l|}{31.36/0.8879} & \multicolumn{2}{l}{29.50/0.8946} \\
    FSRCNN \cite{dong2016accelerating}             &                        & 13       &  6.0      & \multicolumn{2}{l|}{37.00/0.9558} & \multicolumn{2}{l|}{32.63/0.9088} & \multicolumn{2}{l|}{31.53/0.8920} & \multicolumn{2}{l}{29.88/0.9020} \\
    VDSR \cite{kim2016accurate}                    &                        & 666      &  612.6     &\multicolumn{2}{l|}{37.53/0.9587} & \multicolumn{2}{l|}{33.03/0.9124} & \multicolumn{2}{l|}{31.90/0.8960} & \multicolumn{2}{l}{30.76/0.9140} \\
    LapSRN \cite{lai2017deep}                      &                        & 251      &  29.9      & \multicolumn{2}{l|}{37.52/0.9591} & \multicolumn{2}{l|}{32.99/0.9124} & \multicolumn{2}{l|}{31.80/0.8952} & \multicolumn{2}{l}{30.41/0.9103} \\
    DRRN \cite{tai2017image}                       &                        & 298      &  6,796.9   & \multicolumn{2}{l|}{37.74/0.9591} & \multicolumn{2}{l|}{33.23/0.9136} & \multicolumn{2}{l|}{32.05/0.8973} & \multicolumn{2}{l}{31.23/0.9188} \\
    MemNet \cite{tai2017memnet}                    &                        & 678      &   2,662.4  & \multicolumn{2}{l|}{37.78/0.9597} & \multicolumn{2}{l|}{33.28/0.9142} & \multicolumn{2}{l|}{32.08/0.8978} & \multicolumn{2}{l}{31.31/0.9195} \\
    IDN \cite{hui2018fast}                         &                        & 553      &  124.6     & \multicolumn{2}{l|}{37.83/0.9600} & \multicolumn{2}{l|}{33.30/0.9148} & \multicolumn{2}{l|}{32.08/0.8985} & \multicolumn{2}{l}{31.27/0.9196} \\
    CARN \cite{ahn2018fast}                        &                        & 1592     &  222.8     & \multicolumn{2}{l|}{37.76/0.9590} & \multicolumn{2}{l|}{33.52/0.9166} & \multicolumn{2}{l|}{32.09/0.8978} & \multicolumn{2}{l}{31.92/0.9256} \\
    IMDN \cite{hui2019lightweight}                 &                        & 694      &  158.8     & \multicolumn{2}{l|}{38.00/0.9605} & \multicolumn{2}{l|}{33.63/0.9177} & \multicolumn{2}{l|}{32.19/0.8996} & \multicolumn{2}{l}{32.17/0.9283} \\
    PAN \cite{zhao2020efficient}                   &                        & 261      &  70.5      & \multicolumn{2}{l|}{38.00/0.9605} & \multicolumn{2}{l|}{33.59/0.9181} & \multicolumn{2}{l|}{32.18/0.8997} & \multicolumn{2}{l}{32.01/0.9273} \\
    LAPAR-A \cite{li2020lapar}                     &                        & 548      &   171.0    & \multicolumn{2}{l|}{38.01/0.9605} & \multicolumn{2}{l|}{33.62/0.9183} & \multicolumn{2}{l|}{32.19/0.8999} & \multicolumn{2}{l}{32.10/0.9283} \\
    RFDN \cite{RFDN}  &           & 534      &  95.0      & \multicolumn{2}{l|}{38.05/0.9606} & \multicolumn{2}{l|}{33.68/0.9184} & \multicolumn{2}{l|}{32.16/0.8994} & \multicolumn{2}{l}{32.12/0.9278} \\
    RLFN \cite{kong2022residual} &         & 527     &  115.4  & \multicolumn{2}{l|}{38.07/0.9607} & \multicolumn{2}{l|}{33.72/0.9187} & \multicolumn{2}{l|}{32.22/0.9000} & \multicolumn{2}{l}{32.33/0.9299} \\
    BSRN \cite{li2022blueprint}  &         & 332      &  73.0      & \multicolumn{2}{l|}{\textcolor{red}{38.10}/\textcolor{blue}{0.9610}} & \multicolumn{2}{l|}{\textcolor{blue}{33.74/0.9193}} & \multicolumn{2}{l|}{\textcolor{blue}{32.24/0.9006}} & \multicolumn{2}{l}{\textcolor{blue}{32.34/0.9303}} \\
    % VapSR(ours)  &                        & 332      & 74.6      & \multicolumn{2}{l|}{38.07/0.9612} & \multicolumn{2}{l|}{33.76/0.9191} & \multicolumn{2}{l|}{32.27/0.9011} & \multicolumn{2}{l}{32.41/0.9319} \\
    VapSR(ours)  &                        & 329      & 74.0      & \multicolumn{2}{l|}{\textcolor{blue}{38.08}/\textcolor{red}{0.9612}} & \multicolumn{2}{l|}{\textcolor{red}{33.77/0.9195}} & \multicolumn{2}{l|}{\textcolor{red}{32.27/0.9011}} & \multicolumn{2}{l}{\textcolor{red}{32.45/0.9316}} \\ \midrule
    %%%%%%%%%%%%%%%%%%%%%%%%%%%%%%%%%%%%%%%%%%%%%    X2

    Bicubic   & \multirow{16}{*}{$\times 3$}   & -         &  -          & \multicolumn{2}{l|}{30.39/0.8682} & \multicolumn{2}{l|}{27.55/0.7742} & \multicolumn{2}{l|}{27.21/0.7385} & \multicolumn{2}{l}{24.46/0.7349} \\
    SRCNN \cite{dong2014learning}                  &                        & 8        &  52.7      & \multicolumn{2}{l|}{32.75/0.9090} & \multicolumn{2}{l|}{29.30/0.8215} & \multicolumn{2}{l|}{28.41/0.7863} & \multicolumn{2}{l}{26.24/0.7989} \\
    FSRCNN \cite{dong2016accelerating}             &                        & 13       &  5.0       & \multicolumn{2}{l|}{33.18/0.9140} & \multicolumn{2}{l|}{29.37/0.8240} & \multicolumn{2}{l|}{28.53/0.7910} & \multicolumn{2}{l}{26.43/0.8080} \\
    VDSR \cite{kim2016accurate}                    &                        & 666      &  612.6     & \multicolumn{2}{l|}{33.66/0.9213} & \multicolumn{2}{l|}{29.77/0.8314} & \multicolumn{2}{l|}{28.82/0.7976} & \multicolumn{2}{l}{27.14/0.8279} \\
    LapSRN \cite{lai2017deep}                      &                        & 502      &  149.4     & \multicolumn{2}{l|}{33.81/0.9220} & \multicolumn{2}{l|}{29.79/0.8325} & \multicolumn{2}{l|}{28.82/0.7980} & \multicolumn{2}{l}{27.07/0.8275} \\
    DRRN \cite{tai2017image}                       &                        & 298      &  6,796,9   & \multicolumn{2}{l|}{34.03/0.9244} & \multicolumn{2}{l|}{29.96/0.8349} & \multicolumn{2}{l|}{28.95/0.8004} & \multicolumn{2}{l}{27.53/0.8378} \\
    MemNet \cite{tai2017memnet}                    &                        & 678      &  2,662.4   & \multicolumn{2}{l|}{34.09/0.9248} & \multicolumn{2}{l|}{30.00/0.8350} & \multicolumn{2}{l|}{28.96/0.8001} & \multicolumn{2}{l}{27.56/0.8376} \\
    IDN \cite{hui2018fast}                         &                        & 553      &  56.3      & \multicolumn{2}{l|}{34.11/0.9253} & \multicolumn{2}{l|}{29.99/0.8354} & \multicolumn{2}{l|}{28.95/0.8013} & \multicolumn{2}{l}{27.42/0.8359} \\
    CARN \cite{ahn2018fast}                        &                        & 1592     &  118.8     & \multicolumn{2}{l|}{34.29/0.9255} & \multicolumn{2}{l|}{30.29/0.8407} & \multicolumn{2}{l|}{29.06/0.8034} & \multicolumn{2}{l}{28.06/0.8493} \\
    IMDN \cite{hui2019lightweight}                 &                        & 703      &  71.5      & \multicolumn{2}{l|}{34.36/0.9270} & \multicolumn{2}{l|}{30.32/0.8417} & \multicolumn{2}{l|}{29.09/0.8046} & \multicolumn{2}{l}{28.17/0.8519} \\
    PAN \cite{zhao2020efficient}                   &                        & 261      &  39.0      & \multicolumn{2}{l|}{34.40/0.9271} & \multicolumn{2}{l|}{30.36/0.8423} & \multicolumn{2}{l|}{29.11/0.8050} & \multicolumn{2}{l}{28.11/0.8511} \\
    LAPAR-A \cite{li2020lapar}                     &                        & 544      &   114.0    & \multicolumn{2}{l|}{34.36/0.9267} & \multicolumn{2}{l|}{30.34/0.8421} & \multicolumn{2}{l|}{29.11/0.8054} & \multicolumn{2}{l}{28.15/0.8523} \\
    RFDN \cite{RFDN}                    &                        & 541      &  42.2      & \multicolumn{2}{l|}{34.41/0.9273} & \multicolumn{2}{l|}{30.34/0.8420} & \multicolumn{2}{l|}{29.09/0.8050} & \multicolumn{2}{l}{28.21/0.8525} \\
    BSRN \cite{li2022blueprint}                                  &                        & 340      &  33.3      & \multicolumn{2}{l|}{\textcolor{blue}{34.46/0.9277}} & \multicolumn{2}{l|}{\textcolor{blue}{30.47/0.8449}} & \multicolumn{2}{l|}{\textcolor{blue}{29.18/0.8068}} & \multicolumn{2}{l}{\textcolor{blue}{28.39/0.8567}} \\
    % VapSR(ours)  &                        & 339      &  33.8      & \multicolumn{2}{l|}{34.54/0.9284} & \multicolumn{2}{l|}{30.46/0.8446} & \multicolumn{2}{l|}{29.19/0.8078} & \multicolumn{2}{l}{28.42/0.8586} \\
    VapSR(ours)  &                        & 337      &  33.6      & \multicolumn{2}{l|}{\textcolor{red}{34.52/0.9284}} & \multicolumn{2}{l|}{\textcolor{red}{30.53/0.8452}} & \multicolumn{2}{l|}{\textcolor{red}{29.19/0.8077}} & \multicolumn{2}{l}{\textcolor{red}{28.43/0.8583}} \\ \midrule
    %%%%%%%%%%%%%%%%%%%%%%%%%%%%%%%%%%%%%%%%%%%%%%%%%%%%%%%%%   X3

    Bicubic                                       & \multirow{16}{*}{$\times 4$}   & -         &  -          & \multicolumn{2}{l|}{28.42/0.8104} & \multicolumn{2}{l|}{26.00/0.7027} & \multicolumn{2}{l|}{25.96/0.6675} & \multicolumn{2}{l}{23.14/0.6577} \\
    SRCNN \cite{dong2014learning}                  &                        & 8        &  52.7      & \multicolumn{2}{l|}{30.48/0.8626} & \multicolumn{2}{l|}{27.50/0.7513} & \multicolumn{2}{l|}{26.90/0.7101} & \multicolumn{2}{l}{24.52/0.7221} \\
    FSRCNN \cite{dong2016accelerating}             &                        & 13       &  4.6       & \multicolumn{2}{l|}{30.72/0.8660} & \multicolumn{2}{l|}{27.61/0.7550} & \multicolumn{2}{l|}{26.98/0.7150} & \multicolumn{2}{l}{24.62/0.7280} \\
    VDSR \cite{kim2016accurate}                    &                        & 666      &  612.6     & \multicolumn{2}{l|}{31.35/0.8838} & \multicolumn{2}{l|}{28.01/0.7674} & \multicolumn{2}{l|}{27.29/0.7251} & \multicolumn{2}{l}{25.18/0.7524} \\
    LapSRN \cite{lai2017deep}                      &                        & 813      &  149.4     & \multicolumn{2}{l|}{31.54/0.8852} & \multicolumn{2}{l|}{28.09/0.7700} & \multicolumn{2}{l|}{27.32/0.7275} & \multicolumn{2}{l}{25.21/0.7562} \\
    DRRN \cite{tai2017image}                       &                        & 298      &  6,796.9   & \multicolumn{2}{l|}{31.68/0.8888} & \multicolumn{2}{l|}{28.21/0.7720} & \multicolumn{2}{l|}{27.38/0.7284} & \multicolumn{2}{l}{25.44/0.7638} \\
    MemNet \cite{tai2017memnet}                    &                        & 678      &  2,662.4   & \multicolumn{2}{l|}{31.74/0.8893} & \multicolumn{2}{l|}{28.26/0.7723} & \multicolumn{2}{l|}{27.40/0.7281} & \multicolumn{2}{l}{25.50/0.7630} \\
    IDN \cite{hui2018fast}                         &                        & 553      &  32.3      & \multicolumn{2}{l|}{31.82/0.8903} & \multicolumn{2}{l|}{28.25/0.7730} & \multicolumn{2}{l|}{27.41/0.7297} & \multicolumn{2}{l}{25.41/0.7632} \\
    CARN \cite{ahn2018fast}                        &                        & 1592     &  90.9      & \multicolumn{2}{l|}{32.13/0.8937} & \multicolumn{2}{l|}{28.60/0.7806} & \multicolumn{2}{l|}{27.58/0.7349} & \multicolumn{2}{l}{26.07/0.7837} \\
    IMDN \cite{hui2019lightweight}                 &                        & 715      &  40.9      & \multicolumn{2}{l|}{32.21/0.8948} & \multicolumn{2}{l|}{28.58/0.7811} & \multicolumn{2}{l|}{27.56/0.7353} & \multicolumn{2}{l}{26.04/0.7838} \\
    PAN \cite{zhao2020efficient}                   &                        & 272      &   28.2     & \multicolumn{2}{l|}{32.13/0.8948} & \multicolumn{2}{l|}{28.61/0.7822} & \multicolumn{2}{l|}{27.59/0.7363} & \multicolumn{2}{l}{26.11/0.7854} \\
    
    LAPAR-A \cite{li2020lapar}                     &                        & 659      &   94.0     & \multicolumn{2}{l|}{32.15/0.8944} & \multicolumn{2}{l|}{28.61/0.7818} & \multicolumn{2}{l|}{27.61/0.7366} & \multicolumn{2}{l}{26.14/0.7871} \\
    RFDN \cite{RFDN}                &               & 550      &  23.9      & \multicolumn{2}{l|}{32.24/0.8952} & \multicolumn{2}{l|}{28.61/0.7819} & \multicolumn{2}{l|}{27.57/0.7360} & \multicolumn{2}{l}{26.11/0.7858} \\
    RLFN \cite{kong2022residual}         &        & 527     &  29.8  & \multicolumn{2}{l|}{32.24/0.8952} & \multicolumn{2}{l|}{28.62/0.7813} & \multicolumn{2}{l|}{27.60/0.7364} & \multicolumn{2}{l}{26.17/0.7877} \\
    BSRN-S \cite{li2022blueprint}         &        & 156     &  8.3  & \multicolumn{2}{l|}{32.16/0.8949} & \multicolumn{2}{l|}{28.62/0.7823} & \multicolumn{2}{l|}{27.58/0.7365} & \multicolumn{2}{l}{26.08/0.7849} \\
    VapSR-S(ours)              &    & 155      &  9.0     & \multicolumn{2}{l|}{32.14/0.8951} & \multicolumn{2}{l|}{28.64/0.7826} & \multicolumn{2}{l|}{27.60/0.7373} & \multicolumn{2}{l}{26.05/0.7852} \\
    BSRN \cite{li2022blueprint}              &    & 352      &  19.4      & \multicolumn{2}{l|}{\textcolor{blue}{32.35/0.8966}} & \multicolumn{2}{l|}{\textcolor{blue}{28.73/0.7847}} & \multicolumn{2}{l|}{\textcolor{blue}{27.65/0.7387}} & \multicolumn{2}{l}{\textcolor{blue}{26.27/0.7908}} \\
    %old structure
    % VapSR(ours)              &                  & 344      &  19.6 & \multicolumn{2}{l|}{\textcolor{red}{32.35/0.8975}} & \multicolumn{2}{l|}{\textcolor{red}{28.76/0.7850}} & \multicolumn{2}{l|}{\textcolor{red}{27.69/0.7399}} & \multicolumn{2}{l}{\textcolor{red}{26.34/0.7938}} \\ \bottomrule
    %1.1 million iters
    % VapSR(ours)    &   & 342      &  19.5 & \multicolumn{2}{l|}{\textcolor{red}{32.41/0.8982}} & \multicolumn{2}{l|}{\textcolor{red}{28.78/0.7857}} & \multicolumn{2}{l|}{\textcolor{red}{27.69/0.7402}} & \multicolumn{2}{l}{\textcolor{red}{26.40/0.7956}} \\
    %1 million iters
    VapSR(ours)    &   & 342      &  19.5 & \multicolumn{2}{l|}{\textcolor{red}{32.38/0.8978}} & \multicolumn{2}{l|}{\textcolor{red}{28.77/0.7852}} & \multicolumn{2}{l|}{\textcolor{red}{27.68/0.7398}} & \multicolumn{2}{l}{\textcolor{red}{26.35/0.7941}} \\ \bottomrule
\end{tabular}
}
\end{table*}

\subsection{Comparison with State-of-the-art Methods}
We compare the proposed VapSR with exsisting common lightweight SR approaches with upscale factor of $\times2$, $\times3$ and $\times4$, including SRCNN \cite{dong2014learning}, FSRCNN \cite{dong2016accelerating}, VDSR \cite{kim2016accurate}, LapSRN \cite{lai2017deep}, DRRN \cite{tai2017image}, MemNet \cite{tai2017memnet}, IDN \cite{hui2018fast}, CARN \cite{ahn2018fast}, IMDB \cite{hui2019lightweight}, PAN \cite{zhao2020efficient}, LAPAR-A \cite{li2020lapar}, RFDN \cite{RFDN}, RLFN \cite{kong2022residual}, and BSRN \cite{li2022blueprint}.
\tablename~\ref{table:comparison_SOTA} shows the quantitative comparison results for different upscale factors. 
We also provide the number of parameters and Multi-Adds calculated on the $1280\times720$ output. 
Benefit from the simple yet efficient structure, the proposed VapSR achieves state-of-the-art performance with remarkably few parameters. 
Specifically, our VapSR $\times4$ uses 21.68 $\%$ and 28.18 $\%$ parameters of RFDN $\times4$ and IMDN $\times4$, while obtains average 0.187 dB improvement on four evaluation datasets.
Moreover, the proposed VapSR-S achieves competitive performance to BSRN-S \cite{li2022blueprint}, which is the winner of the model complexity sub-track in NTIRE 2022 Challenge on Efficient Super-Resolution \cite{li2022ntire}.
Interestingly, our structure has relative advantages on metric SSIM than PSNR as well.
The results on SSIM can keep on top of the competing models even though the PSNR is slightly lower.

\begin{figure*}[t]
%\newlength-4mm
%\setlength{-4mm}{-0.4cm}
\scriptsize
\centering
\begin{tabular}{ccc}
% 备份
% % one row
\hspace{-0.42cm}
\begin{adjustbox}{valign=t}
\begin{tabular}{c}
\includegraphics[width=0.221\textwidth]{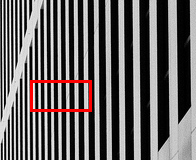}
\\
img\_011 ($\times$4)
\end{tabular}
\end{adjustbox}
\hspace{-2mm}
\begin{adjustbox}{valign=t}
\begin{tabular}{cccccc}
\includegraphics[width=0.149\textwidth]{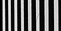} \hspace{-1mm} &
\includegraphics[width=0.149\textwidth]{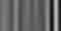} \hspace{-1mm} &
\includegraphics[width=0.149\textwidth]{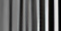} \hspace{-1mm} &
\includegraphics[width=0.149\textwidth]{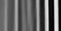} \hspace{-1mm} &
\includegraphics[width=0.149\textwidth]{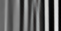} \hspace{-1mm} 
\\
HQ \hspace{-1mm} &
Bicubic \hspace{-1mm} &
IMDN~ \cite{hui2019lightweight} \hspace{-1mm} &
RFDN~ \cite{RFDN} \hspace{-1mm} &
RLFN~ \cite{kong2022residual} \hspace{-1mm}
\\
\includegraphics[width=0.149\textwidth]{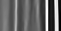} \hspace{-1mm} &
\includegraphics[width=0.149\textwidth]{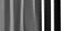} \hspace{-1mm} &
\includegraphics[width=0.149\textwidth]{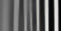} \hspace{-1mm} &
\includegraphics[width=0.149\textwidth]{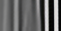} \hspace{-1mm} &
\includegraphics[width=0.149\textwidth]{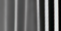} \hspace{-1mm}  
\\ 
VDSR~ \cite{kim2016accurate} \hspace{-1mm} &
IDN~ \cite{hui2018fast} \hspace{-1mm} &
PAN~ \cite{zhao2020efficient} \hspace{-1mm} &
BSRN~ \cite{li2022blueprint}  \hspace{-1mm} &
\textbf{VapSR(ours)} \hspace{-1mm}
\\
\end{tabular}
\end{adjustbox}
\\
% one row
\hspace{-0.45cm}
\begin{adjustbox}{valign=t}
\begin{tabular}{c}
\includegraphics[width=0.221\textwidth]{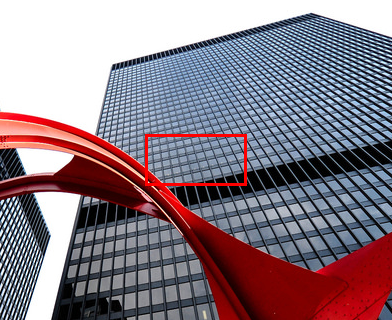}
\\
img\_062 ($\times$4)
\end{tabular}
\end{adjustbox}
\hspace{-2mm}
\begin{adjustbox}{valign=t}
\begin{tabular}{cccccc}
\includegraphics[width=0.149\textwidth]{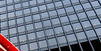} \hspace{-1mm} &
\includegraphics[width=0.149\textwidth]{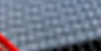} \hspace{-1mm} &
\includegraphics[width=0.149\textwidth]{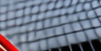} \hspace{-1mm} &
\includegraphics[width=0.149\textwidth]{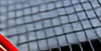} \hspace{-1mm} &
\includegraphics[width=0.149\textwidth]{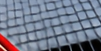} \hspace{-1mm} 
\\
HQ \hspace{-1mm} &
Bicubic \hspace{-1mm} &
IMDN~ \cite{hui2019lightweight} \hspace{-1mm} &
RFDN~ \cite{RFDN} \hspace{-1mm} &
RLFN~ \cite{kong2022residual} \hspace{-1mm}
\\
\includegraphics[width=0.149\textwidth]{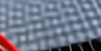} \hspace{-1mm} &
\includegraphics[width=0.149\textwidth]{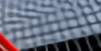} \hspace{-1mm} &
\includegraphics[width=0.149\textwidth]{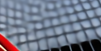} \hspace{-1mm} &
\includegraphics[width=0.149\textwidth]{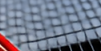} \hspace{-1mm} &
\includegraphics[width=0.149\textwidth]{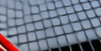} \hspace{-1mm}  
\\ 
VDSR~ \cite{kim2016accurate} \hspace{-1mm} &
IDN~ \cite{hui2018fast} \hspace{-1mm} &
PAN~ \cite{zhao2020efficient} \hspace{-1mm} &
BSRN~ \cite{li2022blueprint}  \hspace{-1mm} &
\textbf{VapSR(ours)} \hspace{-1mm}
\\
\end{tabular}
\end{adjustbox}
\\
% % one row
\hspace{-0.42cm}
\begin{adjustbox}{valign=t}
\begin{tabular}{c}
\includegraphics[width=0.221\textwidth]{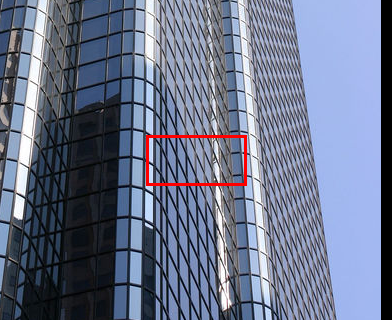}
\\
img\_074 ($\times$4)
\end{tabular}
\end{adjustbox}
\hspace{-2mm}
\begin{adjustbox}{valign=t}
\begin{tabular}{cccccc}
\includegraphics[width=0.149\textwidth]{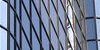} \hspace{-1mm} &
\includegraphics[width=0.149\textwidth]{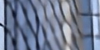} \hspace{-1mm} &
\includegraphics[width=0.149\textwidth]{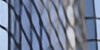} \hspace{-1mm} &
\includegraphics[width=0.149\textwidth]{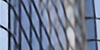} \hspace{-1mm} &
\includegraphics[width=0.149\textwidth]{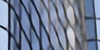} \hspace{-1mm} 
\\
HQ \hspace{-1mm} &
Bicubic \hspace{-1mm} &
IMDN~ \cite{hui2019lightweight} \hspace{-1mm} &
RFDN~ \cite{RFDN} \hspace{-1mm} &
RLFN~ \cite{kong2022residual} \hspace{-1mm}
\\
\includegraphics[width=0.149\textwidth]{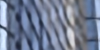} \hspace{-1mm} &
\includegraphics[width=0.149\textwidth]{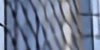} \hspace{-1mm} &
\includegraphics[width=0.149\textwidth]{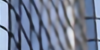} \hspace{-1mm} &
\includegraphics[width=0.149\textwidth]{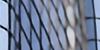} \hspace{-1mm} &
\includegraphics[width=0.149\textwidth]{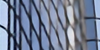} \hspace{-1mm}  
\\ 
VDSR~ \cite{kim2016accurate} \hspace{-1mm} &
IDN~ \cite{hui2018fast} \hspace{-1mm} &
PAN~ \cite{zhao2020efficient} \hspace{-1mm} &
BSRN~ \cite{li2022blueprint}  \hspace{-1mm} &
\textbf{VapSR(ours)} \hspace{-1mm}
\\
\end{tabular}
\end{adjustbox}
\\
% % one row
\hspace{-0.42cm}
\begin{adjustbox}{valign=t}
\begin{tabular}{c}
\includegraphics[width=0.221\textwidth]{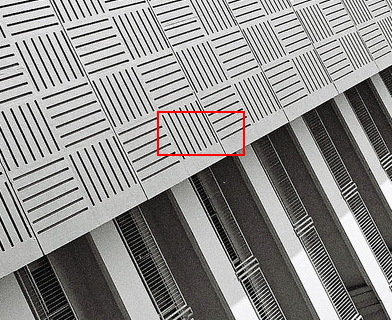}
\\
img\_092 ($\times$4)
\end{tabular}
\end{adjustbox}
\hspace{-2mm}
\begin{adjustbox}{valign=t}
\begin{tabular}{cccccc}
\includegraphics[width=0.149\textwidth]{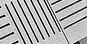} \hspace{-1mm} &
\includegraphics[width=0.149\textwidth]{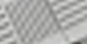} \hspace{-1mm} &
\includegraphics[width=0.149\textwidth]{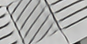} \hspace{-1mm} &
\includegraphics[width=0.149\textwidth]{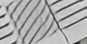} \hspace{-1mm} &
\includegraphics[width=0.149\textwidth]{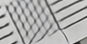} \hspace{-1mm} 
\\
HQ \hspace{-1mm} &
Bicubic \hspace{-1mm} &
IMDN~ \cite{hui2019lightweight} \hspace{-1mm} &
RFDN~ \cite{RFDN} \hspace{-1mm} &
RLFN~ \cite{kong2022residual} \hspace{-1mm}
\\
\includegraphics[width=0.149\textwidth]{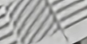} \hspace{-1mm} &
\includegraphics[width=0.149\textwidth]{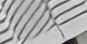} \hspace{-1mm} &
\includegraphics[width=0.149\textwidth]{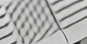} \hspace{-1mm} &
\includegraphics[width=0.149\textwidth]{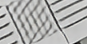} \hspace{-1mm} &
\includegraphics[width=0.149\textwidth]{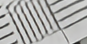} \hspace{-1mm}  
\\ 
VDSR~ \cite{kim2016accurate} \hspace{-1mm} &
IDN~ \cite{hui2018fast} \hspace{-1mm} &
PAN~ \cite{zhao2020efficient} \hspace{-1mm} &
BSRN~ \cite{li2022blueprint}  \hspace{-1mm} &
\textbf{VapSR(ours)} \hspace{-1mm}
\\
\end{tabular}
\end{adjustbox}
\\
\end{tabular}
% \vspace{-3mm}
\caption{Visual comparison about image SR ($\times$4) in some challenging cases.}
\label{Fig:SR}
% \vspace{-4mm}
\end{figure*}

\figurename~\ref{Fig:SR} shows the qualitative comparison of the proposed method.
Our approach can reconstruct stripes and line patterns more accurately than the existing methods, reflecting the advantage of the proposed method on metric SSIM as we mentioned above. 
Take the image ``img\_092'' for example. Most of the existing methods generate noticeable artifacts and blurry effects, while our method produces accurate lines.
For the details of the buildings in ``img\_011'',  ``img\_062'' and ``img\_074'', VapSR could also make reconstruction with fewer artifacts.

% \begin{figure}
%     \centering
%     \includegraphics[width=1\textwidth]{comparison.png}
%     \caption{Visual comparison of VapSR with the state-of-the-art methods on x4 SR.}
%     \label{fig:visual}
% \end{figure}

% \begin{table}[!htbp]
% \caption{Efficient Metrics}
% \end{table}

\subsection{Ablation Study on Micro Design}
\label{sec:ablation}
In this section, we conduct ablation studies on some micro designs involved in our final proposed model.
The micro designs consist of four parts: subordinate components in the architecture, normalization type, attention layers' sequence, and receptive field.
We show their effect on the performance and the best practice of them are all adopted to settle down the eventual structure of VapSR.%, as illustrated in \figurename~\ref{fig:network}.

\begin{table}[!t]
    \centering
    \caption{The accumulated effect of the tricks on subordinate components in the architecture. Here, model (vi) is the same one as in the roadmap, and the experiment on the first row is under the same setting as well. }
    \label{table:micro_design}
    \begin{tabular}{c|ccc|c|c}
        \toprule
        \multirow{2}{*}{Model/Exp. Idx.} & group & inverted & deeper & Parameters & DIV2K\\
        & conv. & bottleneck & up. block & [K] & PSNR \\
         \midrule
        \ model (vi) & - & - & - & 241.1 & 28.95\\
        \ model (vi)+ & $\checkmark$ & - & - & 222.7 & 28.92\\
        \ model (vi)++ & $\checkmark$ & $\checkmark$ & - & 152.2 & 28.84\\
        \ model (vii) & $\checkmark$ & $\checkmark$ & $\checkmark$ & 156.0 & 28.86\\
        \bottomrule
    \end{tabular}
\end{table}

\subsubsection{Subordinate components in architecture.}
We make some detailed designs for the network to achieve better efficiency, as shown in Table. \ref{table:micro_design}. 
These changes are all small but essential tricks and could shed light on designing a high-performance SR network.
The first promoted trick is replacing the regular convolution layer with a group-wise convolution layer at the end of the non-linear mapping module.
We find it a harmless way to reduce the model size while maintaining performance.
The second promoted trick is the inverted bottleneck architecture.
We implement this architecture by expanding and shrinking the channels with the help of the two $1\times1$ convolution layers at both ends of the VAB, as explained in section \ref{sec:methodVAB}.
This trick reduces about 30\% of model parameters while maintaining its PSNR performance in an acceptable region.
The third one is using deeper upsampling layers.
We implement two convolution layers instead of a single one. 
This trick brings performance improvement when introducing acceptable extra parameters.
We apply these tricks to the model (vi) shown in \figurename~\ref{fig:roadmap} step by step, and finally get the model (vii).

\begin{table*}[!tbp]
\centering
\caption{Quantitative comparison of three ablation variables based on model (vii). The performance values surpassing model (vii) are in violet.}
\label{table:ablation}
\resizebox{\linewidth}{!}{
\begin{tabular}{c|l|c|cc|cc|cc|cc}
    \toprule
    \multicolumn{1}{c|}{\multirow{2}{*}{Variable}} & \multicolumn{1}{c|}{\multirow{2}{*}{Method}} & \multirow{2}{*}{Params[K]} & \multicolumn{2}{c|}{Set5} & \multicolumn{2}{c|}{Set14} & \multicolumn{2}{c|}{B100} & \multicolumn{2}{c}{Urban100} \\ %\cline{4-13} 
    \multicolumn{1}{c|}{} & & & \multicolumn{2}{l|}{PSNR/SSIM} & \multicolumn{2}{l|}{PSNR/SSIM} & \multicolumn{2}{l|}{PSNR/SSIM} & \multicolumn{2}{l}{PSNR/SSIM} \\ \midrule

    - & \makecell[l]{model (vii)\{ \\PixelNorm \\5 - 7 - 1 \\k=7, block=10\} } & 156   & \multicolumn{2}{l|}{32.00/0.8929} & \multicolumn{2}{l|}{28.52/0.7797} & \multicolumn{2}{l|}{27.53/0.7343} & \multicolumn{2}{l}{25.79/0.7768} \\ \midrule
    
    \multirow{3}{*}{Normalization} & BatchNorm &  156    & \multicolumn{2}{c|}{-} & \multicolumn{2}{c|}{-} & \multicolumn{2}{c|}{-} & \multicolumn{2}{c}{-} \\
    \multicolumn{1}{c|}{} & InstanceNorm &  156    & \multicolumn{2}{l|}{31.78/0.8897} & \multicolumn{2}{l|}{28.33/0.7764} & \multicolumn{2}{l|}{27.41/0.7328} & \multicolumn{2}{l}{25.15/0.7646} \\
    \multicolumn{1}{c|}{} & GroupNorm  & 156   & \multicolumn{2}{l|}{31.93/0.8912} & \multicolumn{2}{l|}{28.40/0.7766} & \multicolumn{2}{l|}{27.46/0.7323} & \multicolumn{2}{l}{25.62/0.7693} \\ \midrule

    \multirow{1}{*}{Sequence} & 1 - 5 - 7 &  156       & \multicolumn{2}{l|}{32.00/\textcolor{violet}{0.8930}} & \multicolumn{2}{l|}{28.52/\textcolor{violet}{0.7798}} & \multicolumn{2}{l|}{27.53/\textcolor{violet}{0.7347}} & \multicolumn{2}{l}{\textcolor{violet}{25.80/0.7773}} \\
    % wrong value
    % \multicolumn{1}{c|}{} & 1-7-5 & 156    & \multicolumn{2}{l|}{32.00/0.8917} & \multicolumn{2}{l|}{28.44/0.7787} & \multicolumn{2}{l|}{27.48/0.7335} & \multicolumn{2}{l}{25.73/0.7758} \\
    % \multicolumn{1}{c|}{} & 7-5-1 & 156    & \multicolumn{2}{l|}{31.94/0.8912} & \multicolumn{2}{l|}{28.41/0.7781} & \multicolumn{2}{l|}{27.48/0.7335} & \multicolumn{2}{l}{25.71/0.7751} \\ \hline
    \midrule
    
    \multirow{4}{*}{Receptive Field} & k=5, block=11 &  152     & \multicolumn{2}{l|}{\textcolor{violet}{32.01/0.8933}} & \multicolumn{2}{l|}{28.52/\textcolor{violet}{0.7801}} & \multicolumn{2}{l|}{\textcolor{violet}{27.54/0.7349}} & \multicolumn{2}{l}{\textcolor{violet}{25.83/0.7782}} \\
    \multicolumn{1}{c|}{} & k=5, block=12 &  164     & \multicolumn{2}{l|}{\textcolor{violet}{32.02/0.8932}} & \multicolumn{2}{l|}{\textcolor{violet}{28.55/0.7805}} & \multicolumn{2}{l|}{\textcolor{violet}{27.56/0.7356}} & \multicolumn{2}{l}{\textcolor{violet}{25.90/0.7800}} \\
    \multicolumn{1}{c|}{} & k=9, block=9 &  161     & \multicolumn{2}{l|}{31.96/0.8925} & \multicolumn{2}{l|}{28.49/0.7789} & \multicolumn{2}{l|}{27.50/0.7337} & \multicolumn{2}{l}{25.77/0.7761} \\
    \multicolumn{1}{c|}{} & k=11, block=8 & 166      & \multicolumn{2}{l|}{31.93/0.8917} & \multicolumn{2}{l|}{28.47/0.7785} & \multicolumn{2}{l|}{27.49/0.7333} & \multicolumn{2}{l}{25.72/0.7745} \\ \bottomrule
\end{tabular}
}
\end{table*}

\subsubsection{Normalization Types.} 

\begin{figure}[t]
    \subfigure[effect of normalization]{
    \begin{minipage}[t]{0.48\textwidth}
    \centering
        \includegraphics[width=\linewidth]{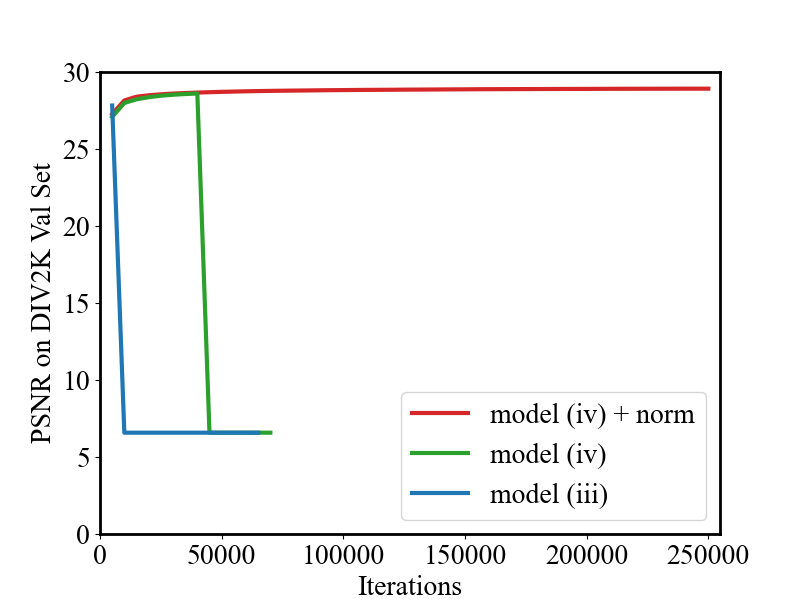}
    \end{minipage}
    }
    \subfigure[comparison of different normalization]{
    \begin{minipage}[t]{0.48\textwidth}
        \centering
        \includegraphics[width=\linewidth]{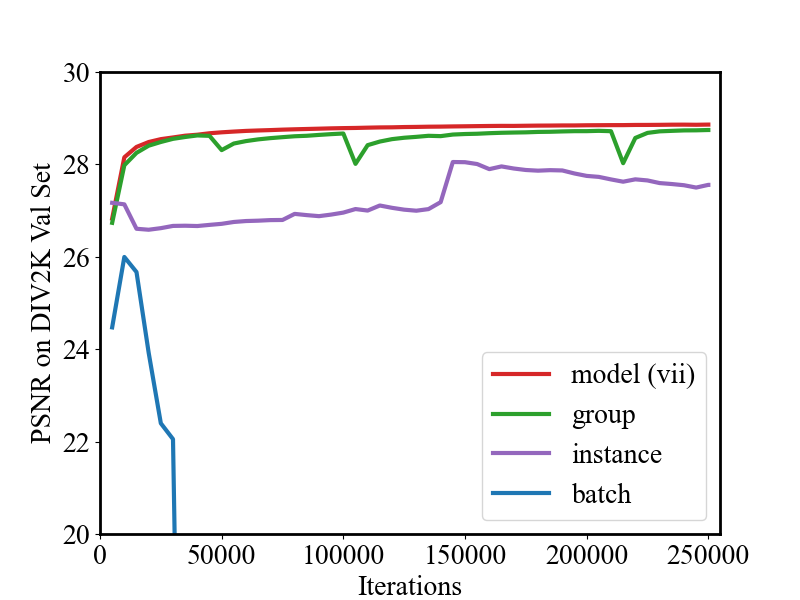}
    \end{minipage}
    }
    \caption{Validation curves with or without normalization under the setting of large learning rate. Plot (a) shows lack of normalization causes the models crash early. In plot (b), the effect of different normalization from good to bad are as followed: pixel normalization (red), group normalization (green), instance normalization (purple) and batch normalization (blue).}
    \label{fig:ways of normalization}
\end{figure}

According to the experiments of designing stage ``stable attention training'', those primary models without any normalization would finally crash at a large learning rate $1\times e^{-3}$.
To demonstrate the effectiveness of the described pixel normalization, we compare it with three common normalization methods, including batch normalization \cite{ioffe2015batch}, instance normalization \cite{ulyanov2016instance} and group normalization \cite{wu2018group}. Since the standard layer normalization \cite{ba2016layer} raises the network's parameters to 2.7M, it's outside the scope of this ablation experiment. The validation curves are shown in \figurename~\ref{fig:ways of normalization}(b) and the results are listed in the Table. \ref{table:ablation}.
It can be observed that batch normalization does not help in preventing training crashes.
Group normalization and instance normalization decrease PSNR and SSIM by a large margin, and the training are not stable or well-converged.
In contrast, our pixel normalization can achieve a steady training curve and the best performance growth.

\subsubsection{The Sequence of Layers in Attention.} 
We conduct the experiment of adjusting the sequence of the attention layers.
\cite{li2022blueprint} proved that putting the pointwise convolution before the depth-wise convolution achieves better performance in the SR task.
Hence, we rearrange the order of the attention layers denoting as $1 \text{ - } 5 \text{ - } 7$, which represents putting $1\times1$ pointwise convolution in front, next the $5\times5$ depth-wise convolution and then $7\times7$ depth-wise dilation convolution.
The experiment show that order $1 \text{ - } 5 \text{ - } 7$ performs slightly better than $5 \text{ - } 7 \text{ - } 1$ which the model (vii) take. This result verifies the conclusion mentioned above again. 
Moreover, we also conducted additional experiments and turns out that the influence of the two depth-wise convolution layers' position is relatively smaller and inconsistent. So we adopted the order $1 \text{ - } 5 \text{ - } 7$ as the final structure.

\subsubsection{Attention Layers' Receptive Field.}
We explored the effect of attention layers with different receptive fields, which are mainly dominated by the dilated convolution layer. 
The field of a $7\times7$ convolution with dilation 3 is equivalent to 19. 
We modify the kernel size of this layer to 5, 9, and 11 separately to change the field and adjust the number of blocks to keep the networks' parameters close to or slightly more than the model (vii)'s. 
As shown in table \ref{table:ablation}, increasing the kernel size to 9 and 11 causes an apparent successive performance drop.
%
% Because a dilation convolution layer of kernel 11 has a field of 31, and the height and width dimensions of the feature sent in the block are just 48 or 64, it's not suitable or worthy to enlarge the kernel continually. 
%
While the smaller kernel size of 5 can obtain performance promotion, even when we keep the parameters less than model(vii)'s with only 11 blocks. 
Therefore we choose the structure with a kernel size of 5 finally.

\section{Conclusions}

This work proposes a lightweight convolutional neural network called VapSR to achieve efficient image super-resolution. The experiments demonstrate that our VapSR can achieve state-of-the-art performance with concise structure and fewer parameters. Starting with the motivation of improving the attention mechanism, we first verified the advantages of using large kernel convolutions on the SR task. Then we successfully applied the efficient depth-wise separable large kernel convolution to reduce the model size. Thirdly, the proposed pixel normalization makes it possible to train this architecture steadily, and we prove its superiority to other normalization methods. We detailed the design process in the form of roadmap. It reveals how we squeeze the complexity of the model while keeping the performance step by step clearly, and lead to VapSR eventually.

\subsubsection{Acknowledgements.} This work is partially supported by the National Natural Science Foundation of China (61906184, U1913210), and the Shanghai Committee of Science and Technology, China (Grant No. 21DZ1100100).

\clearpage

\bibliographystyle{splncs04}
\bibliography{egbib}
\end{document}